\documentclass[aps,pre,twocolumn,floatfix,
nofootinbib,showpacs,longbibliography]{revtex4-1}

\usepackage{mathtools}
\usepackage{color}
\usepackage{color}
\usepackage{blkarray, bigstrut}
\usepackage{relsize}
\usepackage{amsmath}
\usepackage[british]{babel}  
\usepackage[scaled=1.03]{inconsolata} 

\usepackage[colorlinks=true, citecolor=blue, urlcolor=blue]{hyperref}  
\usepackage{graphicx} 
\usepackage[babel]{microtype}  
\usepackage{amsmath,amssymb,amsthm,bm,amsfonts,mathrsfs,bbm} 
\usepackage{xspace}  
\usepackage{pgfplots}
\usepackage{amsmath}
\usepackage{amssymb}

\begin{document}

\title{Structure of passive states and its implication in charging quantum batteries}

\author{Mir Alimuddin}
\email{aliphy80@gmail.com}
\author{Tamal Guha}
\author{Preeti Parashar}
\affiliation{Physics and Applied Mathematics Unit, Indian Statistical Institute, 203 B T Road, Kolkata-700108, India.}


\begin{abstract}
 In this article, in addition to the characterization of geometrical state spaces for the passive states, an operational approach has been introduced to distinguish them on their charging capabilities of a quantum battery. Unlike the thermal states, the structural instability of passive states assures the existence of a natural number $n$, for which $n+1$-copies of the state can charge a quantum battery while $n$-copies cannot. This phenomenon can be presented in a $n$-copy resource-theoretic approach, for which the free states are unable to charge the battery in $n$-copies. Here we have exhibited the single copy scenario explicitly. We also show that general ordering of the passive states on the basis of their charging capabilities is not possible and even the macroscopic entities (viz. energy and entropy) are unable to order them precisely. Interestingly, for some of the passive states, the majorization criterion gives sufficient order to the charging and discharging capabilities. However, the charging capacity for the set of thermal states (for which charging is possible), is directly proportional to their temperature.
\end{abstract}
\maketitle

\section{Introduction} \label{sec1} 
 A state is said to be \textit{passive} if no work (\textit{ergotropy}) can be extracted from it cyclically under unitary evolution\cite{pusz,lenard,allahavardyan}. 
 However, it may be useful to extract work when multiple copies of this state can be accessed globally. Further, if it retains passivity even in the asymptotic limit, then the state is called \textit{completely passive} or \textit{thermal}. These states are the only structurally stable states 
 \cite{lenard,Giovannetti} and take the Gibbsian form associated with an inverse temperature $\beta \geq 0$, whereas for a single passive state, different virtual temperature $\beta_i \geq 0$ can be associated with the different energy levels. But it is not very clear why the passive but not the completely passive states are able to produce work from multiple copies under the reversible (unitary) operations. 
 Intensive research is being conducted in this direction, specially in the fields of work extraction from the non equilibrium quantum states \cite{Giovannetti,marti,carlo,Alhambara}, information theoretic approach in quantum thermodynamics \cite{bera,RUzdin,Brown,FBinder}, passive states energy as entanglement monotone \cite{Mir}, generalised Gibbs states \cite{RBalian,Lostagilo,Halpern,Short,Boes} and achieving Gibbs state from the geometry \cite{DJ}.  
 
 \par
 
  In recent years, various physically motivated ideas have been provided to bring out the distinction between passive and Gibbs states. For example, for every passive state, there exist some copies for which some $\beta_i$ of the composite systems would be negative. This is not the case for a completely passive state, where a unique $\beta \geq 0$ exists, leading to the concept of temperature in the asymptotic limit \cite{skrzypzyckPRE}. In another study \cite{carlo}, a weaker cyclic process has been considered, based on which the passive state's energy can be decreased further and the only states incapable of doing this are in the Gibbsian form. A large dimensional ancilla has been considered as a catalyst to show the energetic instability of the single passive states. An alternative definition of the thermal state is that it is the lowest energetic state under constant entropy or the highest entropic state under constant energy. On the contrary, there exists a unique passive state which is the highest energetic state under constant entropy or the lowest entropic state under constant energy called the {\it maximum energetic} passive state \cite{marti}. In the present work, we have further sharpened this distinction by considering their charging capabilities for a quantum battery.   
 \par
 Geometrically, the set of $d$-dimensional passive states forms a polytope, with $d$ number of vertices. This fact leads us to define the set of witnesses to identify the \textit{non-passive}, i.e., the \textit{active} states.
 
 \par
 
  In our framework for charging quantum batteries (QB), we consider a qubit battery along with a passive or, thermal qudit charger and allow an energy preserving global unitary.
  Quantum batteries were first introduced in \cite{alicki}, followed by many articles \cite{Friis,sai,Farina,Barra,Alicki19,Caravelli} regarding enhancement of charging power \cite{binder,modi,Luis,Dario}, work extraction \cite{Giorgi,Marcello'PRL} and advantage in multiple usage of the battery incorporated with entanglement \cite{Alimuddin,acin}. A battery can be charged up by using the field energy where the unitaries are controlled by an arbitrary field parameter which acts cyclically for finite time. But under these circumstances, a passive battery cannot be charged and so one can consider some additional ancillary systems as a resource for charging. If we take arbitrary ancilla, then it can be charged by just an energy conserving swap operation. This charging procedure has been discussed in \cite{Raam}. However, here we mainly focus on {\it passive/thermal states (restricted resource) as ancilla} since they by themselves are not useful at all. 
  If a charger is not able to charge, then it is called a free state corresponding to the given battery. A condition of free state has been provided considering single copy as ancilla and it is shown that they forms a convex set. Moreover, for every passive state, there exist a positive integer $n$, such that $n$-copies of the state is unable to charge battery, while $(n+1)$-copies can serve the purpose. Hence considering free states with multiple copies, will reduce the set-cardinal and in the asymptotic limit only the thermal states having temperature lower than that of the battery would be free. The reason behind this is the structural instability of passive states for the composite system, which makes them resourceful in multiple copies, unlike the thermal states. Further, we have explicitly provided stochastic matrix for possible battery state transformation under energy conserving global unitary and show that quantum dynamics cannot be advantageous in optimal charging but make some battery state achievable which are unachievable by classical permutations. 
    
 \par
 An ordering among the thermal chargers on the basis of their charging capabilities is possible; the hotter one is the better one. However for the passive states, this kind of ordering is never possible for all $QB$ and even the macroscopic entities like entropy, energy cannot order them precisely. But for some special kind of passive states, the majorization criterion can sufficiently order them. We also provide the activation criterion of a charger for the given battery such that the battery can be made useful in work extraction. 
 Lastly, we focus on the discharging of a battery through the passive state using arbitrary unitary. Again, the majorization criterion provides a sufficient condition for discharging, which is just the opposite of the charging criterion.

\section{Structure of passive states}
Here we will briefly study the set of passive states, along with their possible geometrical representations. A state is said to be \textit{passive}, iff no work can be extracted from it under unitary transformation. Alternatively, the passive states are diagonal in the Hamiltonian basis, with population in each level varying inversely with the energy of that particular level. So a passive state necessarily follows the criterion that (i) $[\rho,H]=0$ and (ii) $\epsilon_i > \epsilon_j$ implies $q_i \leq q_j ~~ \forall i,j$ where, $\epsilon's$ and $q's$ are the eigenvalues of the Hamiltonian and the system respectively. Although no work can be extracted from the single copy of a passive state, one may obtain non-zero work from its multiple copies. However, if no work can be extracted even with infinite copies of a passive state, then the state is said to be \textit{completely passive} or \textit{thermal}. Conversely, an \textit{active} state has potential to extract work unitarily with single copy. However, there is another refinements on the class of passive states, namely \textit{structurally stable}. The stronger condition for structural stability demands that $\epsilon_{i}=\epsilon_{j}\implies q_{i}=q_{j}$. Mathematically, the passive state $\rho$ is said to be structurally stable, {\it iff} there exists a non-increasing function $f$ on the spectrum of the governing Hamiltonian $H$, such that $\rho=f(H)$ which eventually makes their spectrum in the gibbsian form \cite{lenard}. \par 
The set of $d$-dimensional passive states forms a convex polytope in the $(d-1)$-dimensional hyperplane embedded in $d$-dimensional space, where the extreme points are given in Appendix \ref{d-dim}. The convexity of the set of passive states is follows trivially from the definition. However, in the following Lemma we will discuss the status of completely passive, i.e., the thermal states on this polytope.\par 
{\it Lemma 1: There does not exist any thermal state except ${T=0}$ and $T=\infty$ which lies on the boundary of the convex set $\mathcal{S}.$}\\
{\it Proof:} In the $d$ dimensional passive state space, extreme points are represented as $\{e_j=(\underbrace{\frac{1}{j},\frac{1}{j},\frac{1}{j},\cdots,\frac{1}{j}}_\text{j no. of terms},\cdots,0)\}.$ \\
Let a general thermal state of inverse temperature $\beta$ lie on the $(d-1)$ dimensional boundary, which can be constructed by the convex combination of the $(d-1)$ number of extreme points i.e.,\\
$\tau_{\beta}(t_1,\cdots t_d)=\sum\limits_{j=1,j\neq i}^{d}p_je_j $ such that $\sum\limits_{j=1,j\neq i}^{d}p_j=1$,\\

where $i^{th}(t_i)$ and $(i+1)^{th}(t_{i+1})$ element would be equal to $\sum\limits_{j>i}^{d}\frac{p_j}{j}$. \\
So$$\frac{e^{-\beta \epsilon_i}}{z} = \frac{e^{-\beta \epsilon_{i+1}}}{z}. $$\\
Since the Hamiltonian is non-degenerate, $\epsilon_i \neq \epsilon_{i+1}$ and the only solutions are $\beta = 0$, $\infty$. So all other thermal states do not lie on the boundary of the convex set of passive states.$~~~~~~~~~~~~~~~~~~~~~~~~~~~~~~~~~~~~~~~~~~~~~~~~~~~~~~\blacksquare$
\\
Further, being a polytopic structure the set of any $d$-dimensional passive states is compact also. 
The state outside this set $\mathcal{S}$ is called active, useful for work extraction under unitary. Since the state space of the passive state is convex and compact, according to Hahn-Banach theorem for separating hyperplane, it is always possible to construct a set of witnesses to detect these active states.
\par
 \label{theorem 1}
{\bf Theorem 1:} {\it  For any active state $\sigma (\notin S)$, diagonal in energy eigenbasis, (where, $S$ is the set of all passive states for a given Hamiltonian), $\exists$ a Hermitian operator $W$, such that $Tr(W \sigma) < 0$ and $Tr(W \rho) \ge 0, \forall \rho \in S$.}

{\it Proof:}  The passive states in any arbitrary dimension $d$, for a given Hamiltonian, will form a polytope  $\mathcal{P}_{d} \subset \mathbf{R^d}$, which will lie on the $(d-1)$ dimensional hyperplane in $\mathbf{R^d}$ due to the probability constraint. The facets of this polytope will behave as witness operators for the active states diagonal in the energy basis. In general, for the set of $d$-dimensional passive states, there will be $(d+1)$ number of witness operators which are $d\times d$ matrices denoted as, $[W_0, W_{i,(i+1)}, \forall i \in {1,2,...,d}]$. Among these, $W_0$ will be a trivial one, with $[W_0]_{d,d}=1$, and 0 otherwise. 
\\Now, a general $W_{i,(i+1)}$ will be the witness operator with $[W_{i,(i+1)}]_{i,i}=1, [W_{i,(i+1)}]_{(i+1),(i+1)}=-1$ and 0 otherwise $\forall i \in [1,2,...,n]$.$~~~~~~~~~~~~~~~~~~~~~~~~~ ~~~~~~~~~~~~~~~~~~~~~~~~~~~~~~~~~~~~~\blacksquare$

\section{Charging of a quantum battery through Passive state}
\begin{figure}
        \centering
        \includegraphics[width=.40\textwidth]{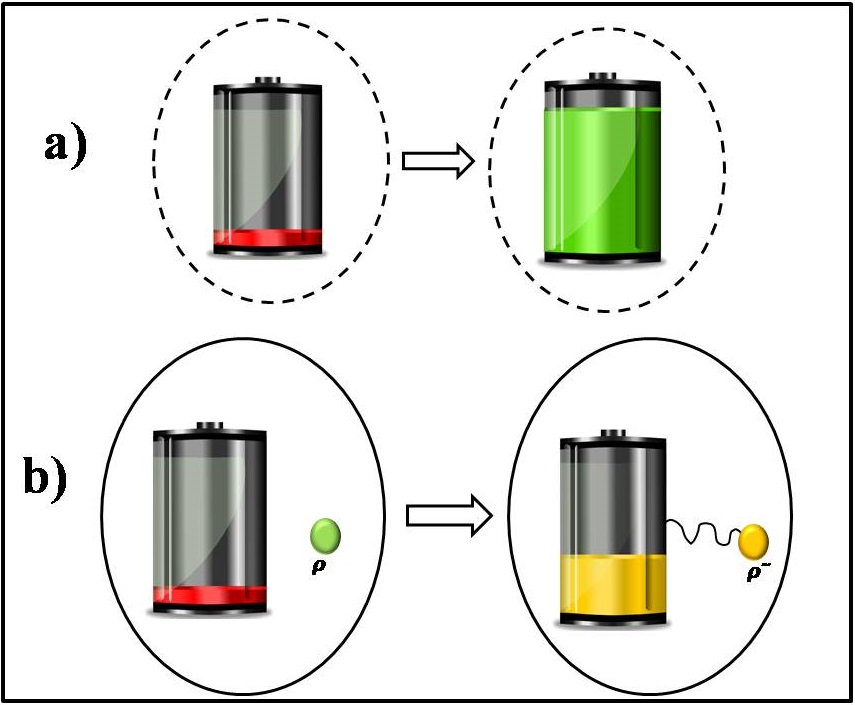}
        \caption{ Charging of a battery under an arbitrary unitary where the corresponding field provides the energy. b) Charging via the anclillary state $\rho$ (passive) where battery and charger form a closed system and evolve under an energy conserving unitary.}
        \label{fig0}
\end{figure}

In general, charging could be done via an arbitrary unitary, where the corresponding field supplies the energy. However, instead we have studied how the energetically passive states could boost up the quantum batteries in the finite dimensional case. To exploit the passive states, we have considered a joint unitary which is energy conserving. A schematic illustration has been given in Figure \ref{fig0}. For simplicity we have taken a completely uncharged battery in the initial state  $|0\rangle_{B}\langle0|$. Although by an arbitrary unitary the battery state can be charged maximally to $|1\rangle_{B}\langle1|$ state, but here the assistance of passive states could impose some restrictions from practical point of view due to energy conserving unitary. Throughout the process we will take the Hamiltonian of the battery as $H_B= |1\rangle\langle1|$ and the Hamiltonian of the passive/thermal state (Charger) as $H_C = \sum\limits_{i=0}^{d-1}i|i\rangle\langle i|$.

\begin{figure}
	\centering
	\includegraphics[width=.50\textwidth]{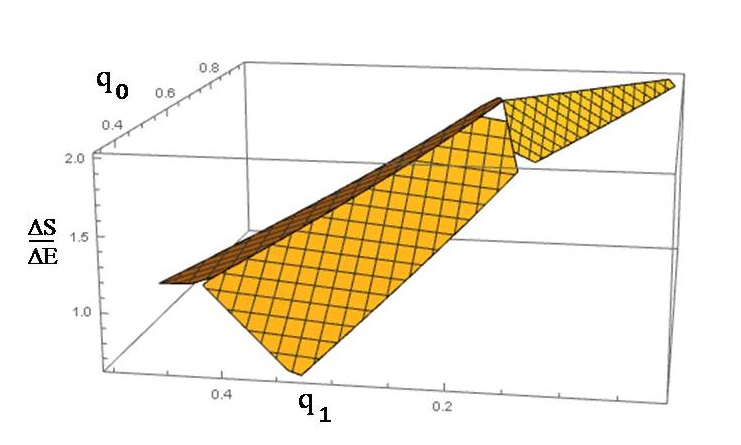}
	\caption{{\it Entropy pollution $\frac{\Delta S}{\Delta E}$ vs passive charger:} Here we have shown how entropy pollution of the given battery $\rho_B = (0.8,0.2)^T$ varies with the passive charger. Though the universal charger $\rho_C = (\frac{1}{3},\frac{1}{3},\frac{1}{3})^T$ makes the entropy pollution lowest, there does not exist any passive charger that can fully swap ($\Delta S = 0$) or supercharge ($\Delta S < 0$) the given battery. The complete mathematical proof has been given in Appendix \ref{stochastic}.}
	\label{charging}
\end{figure}

Let the initial state of the $QB$ be represented by the probability vector $\rho_{B}\equiv(
   1,~
   0)^{T}$ and the $d$ dimensional passive state be given by the probability vector $\rho_C$ $\equiv$ $(
   q_0~
   q_1~
   .~
   .~
   .~
   q_{d-1}
   )^{T}$.The combined initial state is given by $2d \times 2d$ matrix,   
   \begin{equation}
    \rho_{B}\otimes \rho_C = \begin{pmatrix}
    1&0\\
    0 & 0     
    \end{pmatrix} \otimes \begin{pmatrix}
    q_0&0&0\\
    0&q_1&0\\
    0&0&q_2
    \end{pmatrix}
    \end{equation}
  \begin{equation}
     \equiv \begin{pmatrix}
          q_0&0&\cdots&\cdots&\cdots&\cdots&0\\
          0&q_1&\cdots&&&\cdots&\vdots\\
          \vdots & \vdots&\ddots&&&\cdots&\vdots\\
          0&0&\cdots&q_{d-1}&\cdots&\cdots&0\\
          \vdots & \vdots&\cdots&&0&\cdots&0\\
          \vdots & \vdots & \cdots&&&\ddots&\vdots\\
          0 & 0 & \cdots &&&\cdots&0          
      \end{pmatrix}.
      \end{equation}
      For convenience, we will arrange first $d$ number of diagonal elements in column $1$ and the next $d$ number in column $2$ i.e;
   \begin{equation}
   \rho_{B}\otimes \rho_C \equiv \begin{pmatrix}
       q_0&0\\
       q_1&0\\
       q_2&0\\
       \vdots & \vdots\\
       q_{d-1} & 0  
   \end{pmatrix},
   \end{equation}
   where the sum of the columns determine the battery state while sum of the rows give the charger state. Off diagonal elements having the same energy can be interchanged under the energy conserving unitary which is the only allowed unitary operation in this scenario. So the most energetic battery state is given by
   
   $$\tilde{\rho_B} = (q_0,1-q_0)^T.$$
 Now let us consider another charger having state $\rho'_C \equiv (q'_0,q'_1, \cdots, q'_{d-1})^T.$  
 The majorization criterion \cite{R.Bhatia} gives a sufficient condition of a better charger for the given battery state $|0\rangle_{B}\langle0|$ i.e., if $\rho_C \prec \rho'_C$ then $q_0 \leq q'_0$ which implies that the unprimed charger is able to charge more than the primed one. So one can say that a more disordered state is more useful in this scenario. However, with energy entropy of the battery also increases and one cannot extract the whole energy as work. To support this we plot a graph (Figure \ref{charging}) to show how entropy pollution defined by $\frac{\Delta S}{\Delta E}$ changes with the passive states for an arbitrary battery state $\rho_B=(0.8, 0.2)^T$, where $\Delta S$ and $\Delta E$ are the change in entropy and energy of the battery respectively. There does not exist any passive charger that can make $\Delta S \leq 0$. We can see that in case of the battery $\rho_B =(1, 0)^T$, for $q_0 <\frac{1}{2}$ the battery gets activated and the entropy pollution gets lowered.


 \par
   
   If we consider an arbitrary passive $QB$ then the following theorem gives the charging condition on a passive charger.
   \par
   
   \label{theorem 2}
   {\it{\bf Theorem 2:} If the passive $QB$ state is $\rho_B =(p_0,p_1)^T$, then a passive charger $\rho_C=(
      q_0~
      q_1~
      .~
      .~
      .~
      q_{d-1}
      )^{T}$ is able to charge the battery if and only if $\frac{p_0}{p_1} > \min_{i} \{\frac{q_i}{q_{i+1}}\},~~ \forall i \in [0, d-2]$}.\\
{\it Proof:}
The joint state of the battery and the charger is given by
\begin{equation}
    \rho_{B}\otimes \rho_C = \begin{pmatrix}
    p_0&0\\
    0 & p_1     
    \end{pmatrix} \otimes \begin{pmatrix}
    q_0&0&0\\
    0&q_1&0\\
    0&0&q_2
    \end{pmatrix}
    \end{equation}
    \begin{equation}
    \equiv  \begin{pmatrix}
          p_0q_0&0&\cdots&\cdots&\cdots&\cdots&0\\
          0&p_0q_1&\cdots&&&\cdots&\vdots\\
          \vdots & \vdots&\ddots&&&\cdots&\vdots\\
          0&0&\cdots&p_0q_{d-1}&\cdots&\cdots&0\\
          \vdots & \vdots&\cdots&&p_1q_0&\cdots&0\\
          \vdots & \vdots & \cdots&&&\ddots&\vdots\\
          0 & 0 & \cdots &&&\cdots&p_1q_{d-1}          
      \end{pmatrix}.
      \end{equation}
      It can be represented by
\begin{equation}\label{matrix 1}
 \rho_{B} \otimes \rho_C \equiv \begin{pmatrix}
       p_0q_0&p_1q_0\\
       p_0q_1&p_1q_1\\
       p_0q_2&p_1q_2\\
       \vdots & \vdots\\
       p_0q_{d-1} & p_1q_{d-1} 
        \end{pmatrix}.
\end{equation}
Since the pair $p_0q_{k+1}$ and  $p_1q_{k}$, ($\forall k\in[0,d-2]$) are the coefficients of equal energetic states, they can be interchanged by the energy conserving unitary. If any one of the pairs follow $p_0q_{k+1} > p_1q_{k}$, then charging is possible. This leads to the necessary and sufficient condition for charging: $\frac{p_0}{p_1} > \min_{i} \{\frac{q_i}{q_{i+1}}\},~~ \forall i \in [0, d-2]$ $~~~~~~~~~~~~~~~~~~~~~~~\blacksquare$

\par
The probability of a $d$ dimensional passive state can be written as $p_k= e^{\beta_{k+1}}p_{k+1}$, $\forall k \in [0,d-2]$ . For a general passive state, the set $\{\beta_i\}^{d-1}_{i=1}$ can take any positive value without maintaining any particular order.
In case of a completely passive or thermal state a unique virtual temperature $\beta$ can be associated such that $\frac{q_i}{q_{i+1}}=e^{\beta},~~ \forall i \in [0,d-1]$. If the passive battery is define by $\beta_b$ then the charging condition would be $\beta_b > \beta$ which means battery should have lower temperature than the corresponding ancillary thermal state.

\par
 In this work we have used restricted resource i.e; passive or thermal states to aid in the charging of quantum batteries. It is restricted because we have not considered any arbitrary state as ancilla. This makes some transition impossible (full swap/super-charging [\ref{stochastic}]) and it becomes necessary to use active states as ancilla. Apart from distinguishing passive and thermal states, this study is important from the resource theoretic perspective. Here free states are those which cannot be useful in charging where global energy preserving unitary operation is taken as free operation. A criterion has been presented for free states in Theorem 2 and mathematically shown that they form a convex set [\ref{convex}]. But this set of free states does not follow tensor product structure \cite{eric} and may act as resource in multiple copies. If we consider more copies as ancilla, this set of free states would be smaller and in the asymptotic limit, only thermal states of temperature between $0$ and $T$ (temperature of the given battery) would remain as free states [\ref{convex}]. Moreover, we have studied how a battery can be charged in the presence of restricted resource ? What are the possible transitions that can occur ? We have answered these questions and provided a set of stochastic matrix [\ref{stochastic}] for possible battery state transition. Further, we have shown that quantum dynamics can not be advantageous in the optimal charging procedure but makes it possible to achieve many states, which are unachievable by general permuting unitary [\ref{generalstochastic}].

\par  
Here we come back to the main aim of our article (distinguishing passive and thermal states) and investigate whether there exists any order among the passive or among the thermal states in charging quantum batteries.
 Below we provide ordering between the {\it particular type of passive states} on the basis of charging.

\par

{\it {\bf Corollary 1:} An arbitrary passive $QB$ is characterized by inverse temperature $\beta_b$ and the charging states ($\rho_C$ and $\rho'_C$) have been taken such that $\beta_b > \max_i\{\beta_i\}$ and $\beta_b > \max_i\{\beta'_i\}$. So if $\rho'_C \prec \rho_C$ then $\rho'_C$ is a better charger than $\rho_C$.}

{\it Proof:} Since these states are able to charge, they must satisfy Theorem 2, i.e., $\frac{p_0}{p_1} \geq \min_i\{\frac{q_i}{q_{i+1}}\}$.
If a given charging state satisfies $\frac{p_0}{p_1} \geq \max_i\{\frac{q_i}{q_{i+1}}\}$ which means $\beta_b > \max_i\{\beta_i\}$, then all the equal energetic pairs in matrix (\ref{matrix 1}) would swap their positions and the resultant battery state would be given by

\begin{equation}
\tilde{\rho_B}(q)= (p_0-\delta(q),~ p_1+\delta(q)),
\end{equation}
where $\delta(q)=p_0\sum\limits_{i=1}^{d-1}q_i-p_1\sum\limits_{i=0}^{d-2}q_i$.
If $\rho'_C \prec \rho_C $ then $\delta(q') \geq \delta(q)$ which implies $Tr(\tilde{\rho_B}(q')H_B) \geq Tr(\tilde{\rho_B}(q)H_B)$.
$~~~~~~~~~~~~~~~~~~~~~~~~~~~~~~~~~~~~~~~~~~~~~~~~~~~~~~\blacksquare$

\par

{\it {\bf Corollary 2:} A hotter thermal state is a better charger than a colder one.}

{\it Proof:}
For the completely passive or thermal states, all $\beta$ are equal and hence charging of a battery is possible only when the battery state is colder than the charger i.e., $\beta_b > \beta$. Here in the following we have arrange them on the basis of charging capability.
\par
Let us consider a $d$ dimensional $\beta$-thermal charger which can be written as

\begin{equation}
\tau_C= \begin{pmatrix}
       q_0\\
       q_0e^{-\beta}\\
       q_0e^{-2\beta}\\
       \vdots \\
       q_0e^{-(d-1)\beta} 
        
   \end{pmatrix} = \begin{pmatrix}
           q_{d-1}e^{(d-1)\beta}\\
           q_{d-1}e^{(d-2)\beta}\\
          \vdots\\
          q_{d-1}e^{\beta}\\
          q_{d-1} 
           
      \end{pmatrix}
\end{equation}

Probability constraint gives $q_0=\frac{1}{1+x+x^2+\cdots+x^{d-1}}$ and $q_{d-1}=\frac{1}{1+y+y^2+\cdots+y^{d-1}}$ where $y=\frac{1}{x}=e^{\beta}$. If $\beta > \beta' \leftrightarrow y >y' \leftrightarrow x <x'$ gives $q_0 >q'_0$ as well as $q_{d-1}<q'_{d-1}$.

\par

Under interaction, the battery state moves from $\rho_B =(p_0,p_1)$ to $ \tilde{\rho_B}=(p_0-\delta(q),~ p_1+\delta(q))$, where $\delta(q)=(p_0-p_1)+(p_1q_{d-1}-p_0q_0)$. From the above it is clear that the thermal charger having higher temperature, boosts the battery's energy more i.e., $Tr(\tilde{\rho_B}(q)H_B) < Tr(\tilde{\rho_B}(q')H_B)$. $~~~~~~~~~~~~~~~~~~~~~~~~~~~~~~~~~~~~~~~~~~~~~~~~~~~~~~\blacksquare$

\par 

{\bf Ordering of the passive states:} Here in this section, the ordering among the passive states on the basis of their charging capabilities is investigated by explicit example. We have seen that if the battery state is $|0\rangle_B\langle0|$, the charger having lower ground state population ($q_0$) can charge up more. But this parameter alone does not specify complete order. In general, there does not exist any function $f: \mathbb{R}^d\rightarrow \mathbb{R}$ on the charging states which is able to order the passive states on the basis of charging capability for all the battery states simultaneously. Now we provide an example of chargers and batteries for which individual charger is better for the individual battery. Let $\rho_C \equiv (0.5,0.4,0.1)^T$ and $\rho'_C \equiv (0.5,0.3,0.2)^T$. If the battery state is $\rho_B \equiv (0.6,0.4)^T$ then the excited state probability of the battery is increased by $\delta(q)=0.04$ and $\delta(q')=0$, respectively. However, if the battery state is $\rho_B=(0.8,0.2)^T$, then the excited state probability is increased by $\delta(q)=0.22$ and $\delta(q')=0.24$, respectively. From this, we can conclude that there does not exist any function (defined on the charging states only) which can characterize the passive states on the basis of charging capability for {\it all} $QB$ simultaneously.  Moreover, if we are restricted only to the thermal states, then the hotter one is a better charger than the colder one for all $QB$ (of course chargers should have higher temperature than the batteries).\\

\begin{figure}
        \centering
        \includegraphics[width=.40\textwidth]{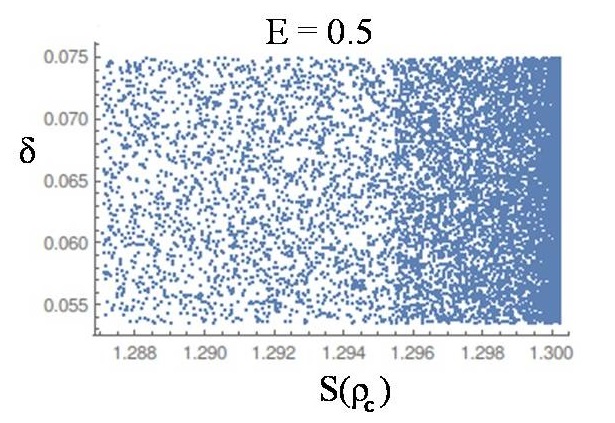}
        \caption{{\it Charging amount $\delta$ Vs. Entropy $S(\rho_C)$:} For a battery state, $\rho_B\equiv(\frac{3}{4},\frac{1}{4})^T$ we have plotted 100,000 random points for the constant energy $E=0.5$. An equilibrium Gibbs state is determined by the macroscopic entities $E$ and $S$ which eventually fix the charging amount $\delta$ of the battery. Whereas a passive (non equilibrium) state cannot be characterized by these two quantities alone e.g. in this figure it has been shown that for constant energy and entropy there exist many states with different charging amount.}
        \label{CE}
\end{figure}

\par
Now we address the question that if a single copy of a charger is unable to charge a $QB$, whether multiple copies can? Such possibilities arise since adding $n$ copies creates more scope to swap between the equal energetic states by using joint unitary $U_{BC_1\cdots C_n} \neq U_B \otimes U_{C_1}\otimes \cdots \otimes U_{C_n} $.
 \par
 
{\it{\bf Corollary 3:} If a thermal state cannot charge a $QB$, then it's multiple copies also cannot.}

\par

{\it Proof:} If the charging is not possible by a thermal state, it means the probability of the battery state satisfies $\frac{p_0}{p_1} \leq \frac{q_0}{q_1}=e^{\beta}$, where $(p_0,p_1)$ is the spectrum of the battery state $\rho_B$ and $\beta$ is the virtual temperature of the corresponding thermal charger $\tau_{C}$. So we will prove that if the single copy of a thermal state cannot charge, it's multiple copies also cannot, i.e.,
\begin{equation}
\begin{aligned}
 & Tr(\rho_B H_B) \nonumber\\
 &=\max_{U}Tr\{Tr_{C}\{U(\rho_B \otimes \tau_{C})U^{\dagger}\}H_B\} \nonumber\\
& = \max_{U}Tr\{Tr_{C}\{U(\rho_B \otimes \tau^{\otimes n}_{C})U^{\dagger}\}H_B\} \nonumber,
\end{aligned}
\end{equation} where, $U$ is energy conserving unitary i.e., $[U,H_{BC}]=0.$

 For a thermal state  the probability of energy $\epsilon_r$ is given by $t_r=t_0e^{-\beta(\epsilon_r-\epsilon_0)}$, where $t_0$ and $\epsilon_0$ is the ground state probability and the corresponding energy respectively. One of the basic features of the thermal state is that the occupying probability for equal energetic eigenstates is equal. Since $\tau_{C}$ is a thermal state, it's $n$ copy also remains thermal at same temperature where probability of the ground state can be defined as $t_0=q^n_0$. Now the probability ratio for the $r$ and $r+1$ energy levels is given by 
\begin{equation}
\frac{t_r}{t_{r+1}}= \frac{t_0e^{-\beta \epsilon_r}}{t_0 e^{-\beta \epsilon_{r+1}}} = e^{\beta},
\end{equation}
which does not satisfy the charging condition. Thus charging is not possible for the given battery $\rho_B$, and even multiple usage of the thermal state cannot enhance the battery energy under joint unitary.  $~~~~~~~~~~~~~~~~~~~~~~~~~~~~~~~~~~~~~~~~~~~~~~~~~~~~~~\blacksquare$

\par

{\it {\bf Theorem 3:} For any passive but not completely passive state $\rho_C$, if $Tr\{Tr_C\{U(\rho_B \otimes \rho_C)U^{\dagger}\}H_B\}=Tr(\rho_BH_B)$} then $ \exists n \in \mathbb{Z}_+ $ s.t. $Tr\{Tr_C\{U(\rho_B \otimes \rho^{\otimes n}_C)U^{\dagger}\}H_B\} = Tr(\rho_BH_B)$ but $Tr\{Tr_C\{U(\rho_B \otimes \rho^{\otimes (n+1)}_C)U^{\dagger}\}H_B\} > Tr(\rho_BH_B)$, where $[U,H_{BC}]=0$. 
 
{\it Proof:} Let us consider a $(d+1)$ dimensional passive state $\rho_C\equiv (q_0,q_1, \cdots,q_d)^T$, which is unable to charge up the passive battery $\rho_B \equiv (p_0,p_1)^T$ i.e.,

$$\frac{p_0}{p_1} < \min_{i} \{\frac{q_i}{q_{i+1}}\} ; ~~~ \forall i \in[0,d-1].$$

If we consider the $(r+1)$ copy of the charging state the probability ratio of $kr$ and $kr+1$ energy levels is given by 
\begin{equation}\label{eq7}
\frac{q_0}{q_1}\times(\frac{q_{k-1}.q_{k+1}}{q^2_k})^{\frac{r}{2}}
\end{equation} 
where $k \in [1,d-1]$. Population ratio of the next consecutive energy levels is given by

\begin{equation}\label{eq8}
\frac{q_1}{q_2} \times (\frac{q^2_k}{q_{k-1}.q_{k+1}})^{\frac{r}{2}}
\end{equation}

Since passive but not completely passive states do not satisfy stability condition, at least for multiple copies some equal energetic states would occur with unequal probabilities. So, if $q^2_k > q_{k-1}.q_{k+1}$, then for some finite values of $r$ Eq.[\ref{eq7}] would satisfy the charging condition i.e.,
$\frac{p_0}{p_1} > \frac{q_0}{q_1}\times(\frac{q_{k-1}.q_{k+1}}{q^2_k})^{\frac{r}{2}}$. Otherwise  
$q^2_k < q_{k-1}.q_{k+1}$ then Eq.[\ref{eq8}] would satisfy the charging condition for some other values of $r$ i.e., $\frac{p_0}{p_1} >\frac{q_1}{q_2} \times (\frac{q^2_k}{q_{k-1}.q_{k+1}})^{\frac{r}{2}}.$

Therefore any passive battery can be charged up by the usage of multiple copies of the passive charger. On the contrary if a single copy of the completely passive (thermal) state cannot charge then it's multiple usage also cannot.$~~~~~~~~~~~~~~~~~~~~~~~~~~~~~~~~~~~~~~~~~~~~~~~~~~~~~~\blacksquare$

\par
\section{Activation of a passive battery}
A passive battery would be called active only when population of the excited state becomes more than the ground state. Then one can extract work from it only through a unitary operation. Even if a passive state has charging capability (by Theorem 2) there is no guarantee that it would make the battery active.

{\it {\bf Theorem 4:}To activate a qubit $QB$, the condition for a 3d charger is given by
\begin{equation}
\frac{p_0}{p_1} < \max \{\frac{1-2q_0}{1-2q_1},\frac{1-2q_1}{1-2q_2},\frac{1-2q_0-2q_1}{1-2q_1-2q_2}\}.
\end{equation}
 }
 
 {\it Proof:} The composite system of battery and charger is given by
 
 \begin{equation}
 \rho_B \otimes \rho_C \equiv
 \begin{pmatrix}
        p_0q_0&p_1q_0\\
        p_0q_1&p_1q_1\\
        p_0q_2&p_1q_2 
         \end{pmatrix}.
 \end{equation}
 The charging conditions are (i)$\frac{p_0}{p_1} > \frac{q_0}{q_1}$ or (ii)$\frac{p_0}{p_1} > \frac{q_1}{q_2}$. If condition (i) is satisfied, then after the action of energy conserving unitary, the probability of the excited state would be
 $$\tilde{p_1}=p_1+(p_0q_1-p_1q_0).$$
 Condition of active state gives
 \begin{equation}
 \begin{aligned}
 p_1+(p_0q_1-p_1q_0) & > \frac{1}{2} \\
\implies \frac{p_0}{p_1} & < \frac{1-2q_0}{1-2q_1}
 \end{aligned}
 \end{equation}
 
 In the same way it can be shown that satisfying condition (ii) gives
 $$\tilde{p_1}=p_1+(p_0q_2-p_1q_1),$$
 and activation implies
 \begin{equation}
 \frac{p_0}{p_1} < \frac{1-2q_1}{1-2q_2}.
 \end{equation}
 
 Simultaneous satisfaction of condition (i) and (ii) makes the state active if 
 \begin{equation}
 \frac{p_0}{p_1} < \frac{1-2q_0-2q_1}{1-2q_1-2q_2}.
 \end{equation}  
 
 From the above equations, a charger would be called an \textit{activator} for the given passive $QB$ iff 
 $$\frac{p_0}{p_1} < \max \{\frac{1-2q_0}{1-2q_1},\frac{1-2q_1}{1-2q_2},\frac{1-2q_0-2q_1}{1-2q_1-2q_2}\}.$$ 
 
 \par
 This can be easily generalized for a charger of any dimension.
$~~~~~~~~~~~~~~~~~~~~~~~~~~~~~~~~~~~~~~~~~~~~~~~~~~~~~~\blacksquare$

\par
Although we have seen that to charge a battery ($\beta_b$) a thermal charger ($\beta$) should be necessarily hotter but it may not sufficiently activate the battery so as to obtain ergotropic work from it. Below we provide a condition on the asymptotic copies of the thermal states (bath) for a given battery such that it can charge as well as activate.
\par
{\bf Theorem 5:} {\it To activate a battery $\rho_B=(p_0,p_1)^T$ a qubit bath given by virtual temperature $(\beta)$ should be upper bounded by $\frac{1}{E}\ln(2p_0)$, where $E$ is the energy difference between the ground and excited state.}

{\it Proof:} Activating a battery through a thermal operation means a battery can freely move from $(p_0,p_1)^T \rightarrow (\frac{1}{2}-\epsilon,\frac{1}{2}+\epsilon)^T$. It will happen if the activated states are thermo-majorized \cite{Horodecki} by the given battery state under the thermal bath $\beta$. Corresponding thermal state is represented by $\tau_{\beta}=(t_0,t_1)$. The battery as well as the thermal ancilla are governed by the same Hamiltonian $H=E|1\rangle \langle 1|$.
\par
\begin{figure}
	\centering
	\includegraphics[width=.40\textwidth]{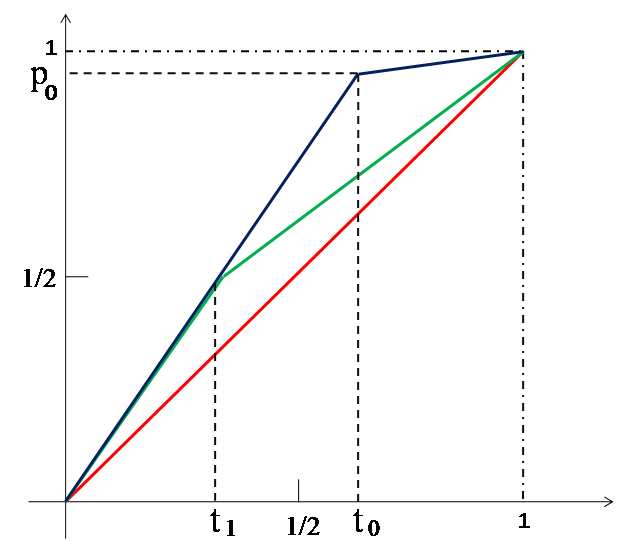}
	\caption{{\it Bound on temperature for activation:} Through this majorization curve we have presented a proof of temperature bound for activation of a given battery $\rho_B = (p_0,p_1)$. The bath state is represented by $\tau_{\beta}=(t_0,t_1)$. For activation, $\rho_B$ should strictly majorize $(\frac{1}{2},\frac{1}{2})$ and lie above in the majorization curve. The battery state would strictly majorize if and only if its initial slope $\frac{p_0}{t_0} > \frac{1}{2t_1}$. To activate the given battery, virtual temperature of the bath ($\beta$) should be strictly upper bounded by $\frac{1}{E}\ln(2p_0)$.}
	\label{Majorization}
\end{figure}

From theorem 2 we know that to charge a battery it must satisfy,
\begin{equation}\label{thermo}
\frac{p_0}{p_1} > \frac{t_0}{t_1} \implies \frac{p_0}{t_0} > \frac{p_1}{t_1}.
\end{equation}
To derive a condition for activation the battery state $(p_0,p_1)$ should strictly thermo-majorize $(\frac{1}{2},\frac{1}{2})$ or thermo-majorize $(\frac{1}{2}-\epsilon,~ \frac{1}{2}+\epsilon)^T$. From Eq. [\ref{thermo}] we can easily show that this would happen if and only if 
\begin{equation*}
\frac{p_0}{t_0} > \frac{1}{2t_1} \implies \beta < \frac{1}{E}\ln(2p_0) < \beta_b = \frac{1}{E}\ln (\frac{p_0}{p_1}).
\end{equation*}$~~~~~~~~~~~~~~~~~~~~~~~~~~~~~~~~~~~~~~~~~~~~~~~~~~~~~~\blacksquare$\\
The graphical proof is depicted in Figure \ref{Majorization}.

\section{Discharging of a quantum battery through the Passive state}

An arbitrary passive $QB$ cannot discharge or be erased via a field unitary only. Consideration of ancilla is necessary to diminish its energy by redistributing energy and entropy further. Here we focus on the discharging process only without taking into account work extraction. The best ancilla is $|0\rangle_{D}\langle0|$ through which any $QB$ can be discharged completely by applying a swap unitary. But how much discharging is possible in the presence of other passive states ? We show below that the ordering among {\it all} passive states in the discharging scenario is exactly inverse to that in the special charging case {\it (Theorem 2 and Corollary 1).}

   {\it {\bf Theorem 6:} A $(d+1)$ dimensional passive discharger $\rho_D$  would discharge a $QB$ if and only if it satisfies $\frac{p_0}{p_1} < \frac{q_0}{q_d}$. Further if $\rho'_D \prec \rho_D$, then $\rho_D$ would be a better discharger than $\rho'_D$.}
  
 {\it Proof:} Let an arbitrary passive $QB$ be denoted by $\rho_B \equiv (p_0,p_1)^T$ and a $(d+1)$ dimensional passive discharger by $\rho_D\equiv (
       q_0~
       q_1~
       .~
       .~
       .~
       q_{d}
       )^{T}.$
 The composite system is given by

   \begin{equation}
   \rho_B \otimes \rho_D \equiv \begin{pmatrix}
          p_0q_0&p_1q_0\\
          p_0q_1&p_1q_1\\
          \vdots&\vdots\\
          p_0q_k  & p_1q_k \\
          p_0q_{k+1} & p_1q_{k+1}\\
          \vdots & \vdots\\
          p_0q_{d-k} & p_1q_{d-k}\\
          \vdots & \vdots \\
          p_0q_{d} & p_1q_{d} 
           \end{pmatrix}.
   \end{equation}
   
   If $p_0q_d < p_1q_0$, then there exists some positive integer $k$ such that
   $$p_1q_{k+1} \leq p_0q_{d-k} < p_1q_k$$ holds, where $k \in [0,\frac{d}{2}]$ when $d$ is even, or $k \in [0,\frac{d+1}{2}]$ when $d$ is odd. So the composite system evolves to (the shifted terms are written in bold)
   \begin{equation}
      U(\rho_B \otimes \rho_D)U^{\dagger} \equiv \begin{pmatrix}
             p_0q_0&{\bf p_0q_{d-k}}\\
             p_0q_1&{\bf p_0q_{d-k+1}}\\
             \vdots&\vdots\\
             p_0q_k  & {\bf p_0q_d} \\
             . & p_1q_{k+1}\\
             \vdots & \vdots\\
             {\bf p_1q_{0}} & p_1q_{d-k}\\
             {\bf p_1q_1}& \vdots \\
             \vdots & \vdots\\
             {\bf p_1q_{k}} & p_1q_{d} 
              \end{pmatrix}
      \end{equation}
      and the final state is given by $\tilde{\rho_B}(q)\equiv (\tilde{p_0},\tilde{p_1})$, where 
      \begin{equation}
      \begin{aligned}
      \tilde{p_0}(q)&=p_0 \sum\limits_{i=0}^{d-k-1}q_i + p_1 \sum\limits_{i=0}^{k}q_i \nonumber \\
      &=p_0(1-\sum\limits_{i=d-k}^{d}q_i)+p_1\sum\limits_{i=0}^{k}q_i \nonumber \\
      &=p_0+p_1\sum\limits_{i=0}^{k}q_i-p_0\sum\limits_{i=d-k}^{d}q_i.
      \end{aligned}
      \end{equation}
      
      If $\rho'_D \prec \rho_D$ then $Tr\{\tilde{\rho_B}(q')H_B\} \geq Tr\{\tilde{\rho_B}(q)H_B\}$ which means that the more ordered state is a better discharger. $~~~~~~~~~~~~~~~~~~~~~~~~~~~~~~~~~~~~~~~~~~~~~~~~~~~~~~\blacksquare$

\section{Conclusion}
 State space of the passive states forms a convex-compact set and becomes simplex for non-degenerate Hamiltonian. We have shown that all the thermal states lie inside the set except $T=0$ and $T=\infty$ which are on the vertex. Any state outside this set is called active. For the diagonal states, we have given finite number of witness operators to detect them.
\par
 We have also discussed how passive states can be useful to charge up the quantum batteries, and provided a criterion for it. Under some additional constraints on the charger states, the majorization criterion is able to order them sufficiently on the basis of their charging capabilities, and the maximally mixed state turns out to be the universal charger for the $QB$. In the case of a thermal charger, the hotter one is always able to charge more than the colder one, with the battery having temperature lower than both. However, there does not exist any such order among the passive states for all the batteries simultaneously and there cannot exist any function defined on the passive charger which can order them on the basis of charging capability. Even the macroscopic entities like energy any entropy cannot order them, unlike in the thermal case. We have provided an explicit example and supported this by graphical illustration. Furthermore, we have shown that if a single copy of a thermal state cannot charge a $QB$, then the asymptotic copies of the same state also cannot. But due to the structural instability of the passive states, any $(n+1)$ copy of the state is able to charge although $n$ copies cannot. The above operational approach is novel way to make their distiction. It says that non equilibrium passive states are always resourceful than the thermal in finite dimensional quatum thermodynamics. We have studied the reverse process, namely, discharging of quantum batteries under an arbitrary global unitary. It turns out that the majorization criterion sufficiently provides order on the discharging capabilities of the passive states.

\par

	Further, these results have also been presented in a resource theoretic framework, for which the free states, which are unable to charge the batteries, form a convex set and the global energy conserving unitary can be considered as free operations. This resource theory does not follow the tensor product structure, as a result, the free states can be super activated with multiple copies. Hence, considering more copies of free states as charger, the cardinality of the set of free states gets smaller. In the asymptotic limit thermal states of temperature $T\leq T_b$ would be the only free states, since all the passive states become resourceful in some copies. In contrast to these passive states of restricted resource, the \textit{active} states are necessary to swap the battery state completely and for super-charging. 
		A set of stochastic matrices is provided to demonstrate all possible transitions of the battery states. Further, we have shown that although quantum dynamics is not advantageous in optimal charging, they achieve some of the classically unachievable states.

\par

It is noteworthy that charging does not sufficiently activate a battery. Hence we have provided a condition that how a single passive state can aid a battery to become active. We have also derived the condition under which a thermal bath can activate a given battery and increase its ergotropy \textit{freely}. It is free because, under this operation the free energy of the initial state always decreases, which in turn decrease the extractable work with respect to the bath although ergotropy of the system increases.

\par

In this article we have considered only qubit battery but one can generalize it for arbitrary dimension. It would also be interesting to investigate explicitly how multiple copies of the charger would effect the charging of a quantum battery precisely. To characterize the minimum number of copies for a passive state required to charge a battery, and whether this problem is decidable or not, may be a future direction for further analysis.    
\section*{Acknowledgment}
M.A. acknowledges financial support from the CSIR project 09/093(0170)/2016-EMR-I.       

\bibliographystyle{unsrt}
\bibliography{bib}

\begin{widetext}
\section{Appendix} \label{Appendix}
\subsection{\textbf{d-dimensional passive states}} \label{d-dim}
   \paragraph*{\textbf{Non-Degenerate Hamiltonian:}}\label{dD}
   A general non-degenerate Hamiltonian can be defined by $H$ = $(
   \epsilon_1~
   \epsilon_2~
   .~
   .~
   .~
   \epsilon_d
   )^{T}$ and a general diagonal state $\rho$ = $(
   p_1~
   p_2~
   .~
   .~
   .~
   p_d
   )^{T}$ with the condition $p_1+p_2+.~.~.~+p_d=1$. By the definition of passive state, $p_1 \geq p_2 \geq .~.~.~\geq p_d$ if $\epsilon_1 \leq \epsilon_2 \leq .~.~.\leq \epsilon_d$.
   \par The passive state space would occur in $(d-1)$ dimensional space due to the normalization constraint.  There would be $d$ number of extreme points, among which two are thermal ($T=0$ and $T=\infty$) and other ($d-2$) are passive states. \\
   Extreme points:  $e_1(T=0)$ = $(
   1 ~~
   0 \cdots
   0
   )^{T}$, $e_2$ = $(
   \frac{1}{2}~~
   \frac{1}{2}~~
   0\cdots
   0
   )^{T}$, $e_3$ = $(
   \frac{1}{3}~~
   \frac{1}{3}~~
   \frac{1}{3}\cdots
   0
   )^{T}$, $\cdots, e_i = (\overbrace{\hbox{$\frac{1}{i} \cdots \frac{1}{i}$}}^{\hbox{i}},~0~0\cdots~0)^T$ .~.~,
   $e_d(T=\infty)$ = $(
   \frac{1}{d}~~
   \frac{1}{d}~~
   \frac{1}{d} \cdots
   \frac{1}{d}
   )^{T}$. 
   \subsection{Free states of battery form convex set}\label{convex}  
   If a charger cannot charge a given battery then we call it free state or free charger of that battery. Although by definition this set of free states follows convexity but here we have shown it by simple mathematics. From theorem 2 we have the charging condition $\frac{p_0}{p_1} > \min_i\{\frac{q_i}{q_{i+1}}\}
    $, where $\{p_i\}$ and $\{q_i\}$ are the probabilities of battery and charger state. Let we conside another free state by $\{q'_i\}$ and for convexity we need to show that
    \begin{equation*}
    \frac{p_0}{p_1} \leq \min_i\{\frac{r_i}{r_{i+1}}\}
    \end{equation*}
    where $r_i = \lambda q_i + (1-\lambda) q'_i$ for $0 \leq \lambda \leq 1$. 
    Let us prove it by negation. If
    \begin{equation}
    \begin{aligned}
    &~~~~~~\frac{p_0}{p_1} > \frac{\lambda q_i + (1-\lambda) q'_i}{\lambda q_{i+1} + (1-\lambda) q'_{i+1}} \nonumber \\
    & \implies \lambda p_0q_{i+1} + (1-\lambda) p_0q'_{i+1} > \lambda p_1q_{i} + (1-\lambda) p_1q'_{i} \nonumber \\
    & \implies (p_1q'_i-p_0q'_{i+1}) + \lambda (p_1q_i-p_0q_{i+1})+\lambda(p_0q'_{i+1}+p_1q'_i) < 0.
    \end{aligned}
    \end{equation}
    Following charging conditions all the parentheses are positive. Since $\lambda$ is positive this cannot be true. Therefore convex combination of two arbitrary free charger is also free, i.e; 
    \begin{equation*}
    \frac{p_0}{p_1} \leq \min_i \{\frac{r_i}{r_{i+1}}\}.
    \end{equation*}
    \par
    We have taken only a single charger state as an ancilla and provided its charging condition and the set of free states. By definition, a passive state can become active in multiple copies and eventually able to  charge up a given battery. So it is obvious that the free states do not follow tensor product structure. As we increase the copies of the ancilla, the set would become smaller. If we define the set of $n$ copy free states for a given battery by $\mathcal{F}_n$ then $\mathcal{F}_1 \supset \cdots \supset \mathcal{F}_n\supset \cdots \supset \mathcal{F}_{\infty}$. We have shown that asymptotically every passive state becomes a resource and can charge any battery (Theorem 3). On the other hand the set of thermal states in $\mathcal{F}_1$ and $\mathcal{F}_{\infty}$ are always similar, because, according to Corollary 3 if a single thermal states cannot charge a QB then its multiple copies also cannot. Thermal states having temperature ($T$) lower than battery ($T_b$) are always considered as free (Corollary 2) i.e; $0 \leq T \leq T_b$. So a bath containing asymptotic copies of thermal states of temperature $T\leq T_b$ can never be useful in charging a qubit battery of $T_b$.

   \subsection{Stochastic matrix}   \label{stochastic}
  Any transition on the probability vector can be visualized by a stochastic matrix. Here we have provided the set of stochastic matrices for battery state transition. A battery state can be transformed in the given way,
  \begin{equation}
  \tilde{\rho_B} = Tr_C \{U(\rho_B \otimes \rho_C)U^{\dagger}\}, 
  \end{equation}     
   where $\rho_C=(q_0,q_1,\cdots,q_{d})$ can be any diagonal state not necessarily passive, $U$ is energy conserving on the joint system. This is a completely positive trace preserving (CPTP) map. If we restrict to the optimal charging procedure, unitary should be restricted to the permutation on the equal energy spaces. First we will give the stochastic matrix for permuting cases only, the general case follows in the next section. Let us make the transition $|0k\rangle \leftrightarrow |1k-1\rangle$ and the transformed battery state is represented by $\tilde{\rho_B} = S^k (\rho_B)$, where $S^k$ is the stochastic matrix defined on $k$ th energy permutation. The following transformation is expressed by  
   \begin{equation}
\tilde{\rho_B}=S^k (\rho_B)=\begin{pmatrix}
    1-q_k&q_{k-1}\\
    q_k&1-q_{k-1} 
\end{pmatrix} \begin{pmatrix}
    p_0\\
    p_1 
\end{pmatrix} =  \begin{pmatrix}
    p_0-\delta_k\\
    p_1+\delta_k 
\end{pmatrix}   
      \end{equation} 
      where $\delta_{k}=p_0q_k-p_1q_{k-1}$.
    It can be extended straight forwardly for multiple permutations, for example if we permute $k$ th ($|0k\rangle \leftrightarrow |1k-1\rangle$) and $m$ ($|0m\rangle \leftrightarrow |1m-1\rangle$) th energy simultaneously, the stochastic matrix representation would be following,
      \begin{equation}
      \tilde{\rho_B}=S^{k,m} (\rho_B)=\begin{pmatrix}
          1-q_k-q_m&q_{k-1}+q_{m-1}\\
          q_k+q_m&1-q_{k-1}-q_{m-1} 
      \end{pmatrix} \begin{pmatrix}
          p_0\\
          p_1 
      \end{pmatrix} =  \begin{pmatrix}
          p_0-\delta_k-\delta_m\\
          p_1+\delta_k+\delta_m 
      \end{pmatrix}   
            \end{equation}        
  If we restrict to the passive ancilla then allowed stochastic matrices would be constrained by their matrix component $S_{01}^{(\cdots)} \geq S_{10}^{(\cdots)}$ i.e; $q_{k-1}+q_{m-1} \geq q_{k}+q_{m}  $. It assures that active states are necessary for full swap and super-charging. Below we have shown this in details.
            
            \par
         
           {\it Full swapping and super-charging:} Full swapping means inverting population which keeps the entropy unchanged but increases energy. If the entropy decreases with increasing energy it is called super-charging. Below we provide the condition on stochastic matrices when full swap on the battery state and super charging is not possible. Let,           
           \begin{equation}
           \begin{pmatrix}
               a&b\\
               1-a&1-b 
           \end{pmatrix}\begin{pmatrix}
                          p_0\\
                          p_1 
                      \end{pmatrix} = \begin{pmatrix}
                                     ap_0+bp_1\\
                                     (1-a)p_0+(1-b)p_1 
                                 \end{pmatrix}  
           \end{equation}            
        Condition for inverted population,         
        \begin{eqnarray*}
        ap_0+bp_1=p_1 \\
        \frac{a}{1-b}=\frac{p_1}{1-p_1}
        \end{eqnarray*}
        Since the battery state is passive i.e; $p_0 \geq p_1$, the necessary condition of full swap is that $b\leq(1-a)$. If we add a passive state as ancilla, this condition would never be satisfied, so one must allow some active state. To decrease entropy further while energy keeps increasing it can be easily shown that, $\frac{a}{1-b} < \frac{p_1}{p_0}$, which again demands that the ancilla should be an active state.
       
  \subsection{Quantum dynamics} \label{generalstochastic}     
  In order to charge, the equal energetic states of the joint system can be either completely permuted or transformed into a coherent superposition by some energy conserving unitary. Here the former is classical whereas the later is truly quantum dynamics. We shall show that in the optimal charging procedure quantum signature is absent. Consider two equal energetic states $|0k\rangle$ and $|1k-1\rangle$ which can be transfered by general unitary as follows, 
  \begin{eqnarray}\label{cohunitary}
  U|0k\rangle = sin\alpha_k |0k\rangle + icos\alpha_k|1k-1\rangle \nonumber \\
  U|1k-1\rangle = icos\alpha_k |0k\rangle + sin\alpha_k|1k-1\rangle.
  \end{eqnarray}                 
    The occupation of the joint state is changed to,    
    \begin{eqnarray}\label{cohprobability}
    \tilde{p_0q_k} = sin^2(\alpha_k) p_0q_k + cos^2(\alpha_k)p_1q_{k-1} \nonumber \\
    \tilde{p_1q_{k-1}} = cos^2(\alpha_k)p_0q_k + sin^2(\alpha_k)p_1q_{k-1}.
    \end{eqnarray}   
      If $p_0q_k \geq p_1q_{k-1}$ then optimal charging would occur at $\alpha_k=0$ (complete permutation) such that $\tilde{p_0q_k}=p_1q_{k-1}$ and $\tilde{p_1q_{k-1}}=p_0q_k$. If $p_0q_k \leq p_1q_{k-1}$; then optimal charging would occur at $\alpha_k=\frac{\pi}{2}$ (identity operation) such that $\tilde{p_0q_k}=p_0q_{k}$ and $\tilde{p_1q_{k-1}}=p_1q_{k-1}$. So generating coherence in energy basis cannot help in optimal charging.  
   
   \par  
  From Eq.[\ref{cohunitary}] and Eq. [\ref{cohprobability}], the general stochastic matrix can be written as
  \begin{equation}
  \tilde{S^k}= \begin{pmatrix}
      1-q_kcos^2(\alpha_k)&q_{k-1}cos^2(\alpha_k)\\
      q_kcos^2(\alpha_k)&1-q_{k-1}cos^2(\alpha_k) 
  \end{pmatrix}
  \end{equation}
  Simultaneous considertaion of other equal energetic states i.e; $|0m\rangle$ and $|1m-1\rangle$ and creating coherence between them by $\alpha_m$ just like Eq.[\ref{cohunitary}] is given by the stochastic matrix
  \begin{eqnarray}
  \tilde{S}^{k,m}= \begin{pmatrix}
        1-q_kcos^2(\alpha_k)-q_mcos^2(\alpha_m)&q_{k-1}cos^2(\alpha_k)+q_{m-1}cos^2(\alpha_m)\\
        q_kcos^2(\alpha_k)+q_mcos^2(\alpha_m)&1-q_{k-1}cos^2(\alpha_k)-q_{m-1}cos^2(\alpha_m)
    \end{pmatrix} 
  \end{eqnarray} 
    If we consider more equal energetic states and change them unitarily, in the stochastic matrix those terms should be added just the same way. $\alpha \in \{0,1\}$ is the class of stochastic matrix that is equivalent with permutation and identity operation, others values of $\alpha$ are purely quantum which takes the initial battery states to many unachievable states.
\end{widetext}
\end{document}